\documentclass[12pt,preprint]{aastex}

\usepackage{psfig}

\advance \voffset by 0.00cm\relax

\begin{document}

\title{MERGER SITES OF DOUBLE NEUTRON STARS AND THEIR HOST GALAXIES}

 \author{KRZYSZTOF BELCZYNSKI\altaffilmark{1,2,3}, TOMASZ BULIK\altaffilmark{2} 
         AND VASSILIKI KALOGERA\altaffilmark{1}}

 \affil{
     $^{1}$ Northwestern University, Dept. of Physics \& Astronomy,
       2145 Sheridan Rd., Evanston, IL 60208\\
     $^{2}$ Nicolaus Copernicus Astronomical Center,
       Bartycka 18, 00-716 Warszawa, Poland;\\   
     $^{3}$ Lindheimer Postdoctoral Fellow\\  
     belczynski@northwestern.edu, bulik@camk.edu.pl,
     vicky@northwestern.edu}

 \begin{abstract} 
Using the {\em StarTrack} population synthesis code we analyze the 
formation channels possibly available to double neutron star binaries and
find that they can be richer than previously thought. We identify a group
of short lived, tight binaries, which do not live long enough to escape
their host galaxies, despite their large center-of-mass velocities. We
present our most recent results on all possible evolutionary paths leading
to the formation of double neutron stars, calculate their coalescence
rates, and also revisit the question of the distribution of merger sites
around host galaxies. For a wide variety of binary evolution models and
galaxy potentials, we find that most of neutron star mergers take place
within galaxies.  Our results stem from allowing for radial and common
envelope evolution of helium-rich stars (testable in the future with
detailed stellar-structure and hydrodynamic calculations) and indicate
that double neutron star binaries may not be excluded as Gamma-Ray Burst
(GRB) progenitors solely on the basis of their spatial distribution 
around host galaxies.
We also find, in contrast to Bethe \& Brown (1998), that in a significant
fraction of common envelope (CE) phases neutron stars do not accrete 
enough material to become black holes, and thus the channels involving 
CEs are still open for the formation of double neutron stars. 
 
 \end{abstract}


\section{INTRODUCTION}

Thirty years after the discovery of GRBs,
their distance scale has been firmly established with the measurement of
redshifts (Metzger et al. 1997). Cosmological GRB scenarios require an energy
release of about $10^{51}$\,ergs,   pointing at violent events in the stellar
evolution scenarios. Coalescences of double neutron star (NS-NS) binaries have
been considered as possible GRB source (Paczynski 1986; Eichler et al. 1989).   
On theoretical grounds the timescale of such a merger is
expected to be short (e.g., Ruffert \& Janka 1999), of
the order of a few milliseconds,  however numerical simulations show 
that they might last up to $0.5$\,s  (e.g. Lee \& Kluzniak 1998).
The discovery and observations of GRB afterglows and the identification of host
galaxies have allowed comparisons of theoretical distributions of merger sites
with the observed distribution of afterglow positions relative to GRB host
galaxies. Such calculations (Bloom, Sigurdsson, \& Pols 1999; Bulik,
Belczynski, \& Zbijewski 1999; Fryer, Woosley, \& Hartmann 1999; Bloom,
Kulkarni, \& Djorgovski 2002)  have found that NS-NS systems acquire rather
high velocities and have long enough lifetimes that a large fraction of the
coalescence events takes place outside the host galaxies. Black-hole
neutron-star (BH-NS) mergers have also been found to take place outside host
galaxies, yet with the tighter distribution of their merger sites.

In this paper we present a thorough investigation of the formation paths 
and merger sites of binaries containing two neutron stars using the 
{\em StarTrack} population synthesis code. 
In particular, we explore the implications of allowing for CE episodes 
involving helium stars.

\section{MODEL DESCRIPTION}
 
In our calculations we use the {\em StarTrack} population synthesis code
described in detail in Belczynski , Kalogera \& Bulik (2002a).  
Dynamical evolution of binaries, in particular effects of kicks and mass 
loss due to supernovae (SN) explosions, in diffrent galactic potentials is 
presented in Belczynski, Bulik \& Rudak (2002b). 

In order to investigate the systematic inaccuracies of binary population 
synthesis methods, we vary many of the parameters describing the stellar 
evolution. 
We define standard model with: 
Cordes \& Chernoff (1998) kick velocity distribution; 
maximum NS mass $M_{\rm max,NS}=3.0$\,M$_\odot$;
$\alpha_{\rm CE}\times\lambda = 1$, where $\alpha_{\rm CE}$ -- 
CE efficiency (Webbink 1984), 
$\lambda$ -- numerical factor describing stellar density 
distribution (de Kool 1990);
f$_{\rm a}=0.5$, $j_{\rm a}=1.0$, in non-conservative stable MT, 
f$_{\rm a}$ denotes the part of the mass lost by donor and accreted 
by the companion and the rest is lost from a system with angular 
momentum equal to ${2 \pi j_{\rm a} A^2 / P}$, where $A$ and $P$ are 
the orbital separation and period, respectively; 
$M_{\rm conv}=4.5$$\,M_\odot$, evolved helium stars of mass below 
$M_{\rm conv}$ are assumed to develop deep convective envelopes;
continuous star formation rate (SFR);
initial mass function (IMF) $\propto M_1^{-2.7}$;
$f_{\rm bi}=50 \%$, binary fraction;
$\Phi(q)=1$, initial mass ratio distribution, $q \equiv M_2/M_1$, where
$M_1$ is mass of primary, and $M_2$ is mass of secondary;  
partial fall back (FB) allowed for stars with $5.0 < M_{\rm CO} < 7.6 \,M_\odot$, 
where $M_{\rm CO}$ is final stellar CO core mass;
in CEs hyper-critical accretion (HCA) onto NS/BH allowed.
All the other models, presented in Table~\ref{rates}, differ only by the value of 
one parameter from the standard model (A) and are described in detail in 
Belczynski et al. (2002b).

\section{RESULTS}

Our results depend crucially on the treatment of immediate progenitors 
of neutron stars (NS): low-mass  helium stars, and in particular their 
radial expansion as well as their behavior during mass transfer (MT) 
episodes in close binaries.
Although the radial expansion has been acknowledged in the literature,
it has not been widely used in population synthesis studies.
In our earlier work (Belczynski \& Kalogera 2001) we argued that the 
two evolved low-mass helium stars (and thus with at least partially 
convective envelopes) may drive dynamically unstable MT, leading to CE 
evolution.
In the present study we allow low-mass 
helium stars to expand during their evolution, 
and possibly initiate MT in close binaries.
If an evolved low-mass helium star is more massive than its companion, we 
assume that the MT will be dynamically unstable and will lead to CE phase.
We treat such CE phases with the standard ``alpha'' prescription (Webbink 1984), 
to check whether systems survive the episode or result in mergers.
The standard description of CE phase has been used successfully for H-rich giants, 
however without detailed hydrodynamical simulations, we cannot address the
question whether it also works in the case of low-mass He-rich giants.  
Since hydrodynamical simulations are beyond the scope of this study, we also 
present calculations in which we assume that all MT episodes of low-mass
helium stars lead to mergers, and in which the helium stars are not allowed 
to expand radially.

\subsection{Formation Channels}

We find that coalescing NS-NS binaries are formed in various ways, including
more than 14 different evolutionary channels (Belczynski et al. 2002a). 
However, the entire population  may be divided into three groups.
{\em Group I} consists of non-recycled NS-NS systems identified
by  Belczynski \& Kalogera (2001). Progenitors of these systems 
end their evolution in a double  CE of two evolved helium
stars.  If a merger is avoided and the CE is ejected, a tight
binary  consisting of two bare carbon-oxygen (CO) cores is
formed.  The CO cores form neutron stars in two consecutive Type
Ic SN explosions.  Provided, that the system is not disrupted by
SN kicks and mass loss, two  neutron stars form an eccentric and
tight binary with the unique  characteristic that none of the
neutron stars had a chance to be recycled.  Progenitors of {\em
Group II} NS-NS binaries end their evolution  in a CE episode
involving an evolved low-mass helium star donor and a neutron
star companion.  Despite of the short timescale of the CE
phase, the NS may increase its mass due  to hyper-critical
accretion (e.g., Brown 1995;  Bethe \& Brown 1998). If
system avoids merger in the CE phase and if NS does not collapse
into a black hole (BH), a system consisting of
recycled pulsar and bare CO core is formed. After a Type Ic SN
explosion, a very tight and eccentric NS-NS binary  is formed. 
These systems (see also Belczynski et al.\ 2002a) are
characterized by  very short lifetimes. Systems of {\em Group
III}, consisting of all other NS-NS, are  formed along more or
less classical channels (Bhattacharya \& van den Heuvel  1991).

\begin{figure*}[t]
\centerline{\psfig{file=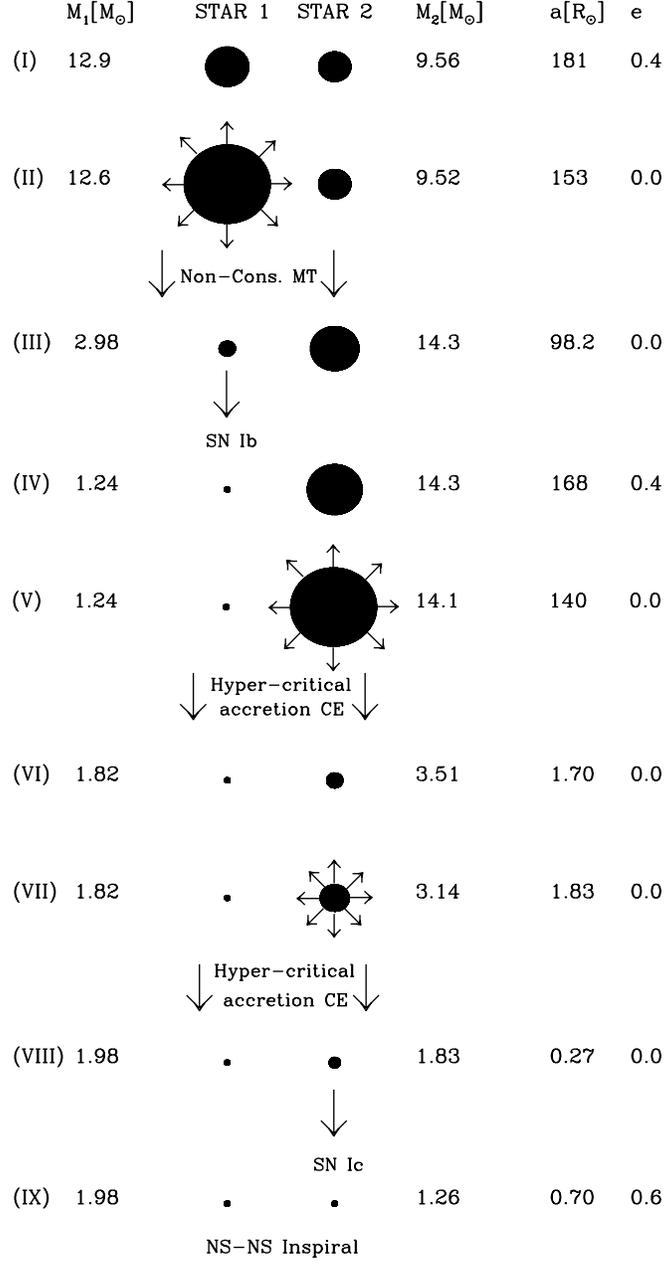,height=0.8\textheight}}
\caption{
Stages of the dominating NS-NS formation path; 
after 23 Myr of evolution and two SN, a tight NS-NS binary with 
a merger time of about 0.7 Myr is formed (details are given in 
\S\,3.1).
}
\label{NSevol}
\end{figure*}

In Figure~\ref{NSevol} we show an example of the formation of a tight
NS-NS binary of Group II. The evolution begins with two massive
stars in a rather wide and eccentric orbit (stage I). The primary evolves
off the main sequence, expands to giant dimensions circularizing
the initially eccentric orbit. Once the primary
reaches its Roche lobe, nonconservative but dynamically stable MT 
begins (stage II). The donor envelope is in part accreted
onto the main sequence (MS) companion and in part lost from the binary. 
The post-MT binary
(stage III) consists of the exposed helium core of the primary and the MS
secondary, which increased its mass and was rejuvenated.
The low-mass helium star evolves through core and shell helium burning,
and radially expands significantly, although it never fills its Roche lobe. 
Its evolution ends with a Type Ib SN explosion forming
the first NS in the system. Due to the asymmetric SN explosion and the
associated mass loss, the post-SN system is eccentric and the orbit widens
(stage IV). Next the secondary ends core hydrogen burning and evolves
toward the giant branch. The associated radial expansion leads to Roche
lobe overflow, and the orbit is again tidally circularized.  Since in this
case the donor is much more massive than its NS companion, the issuing MT
is dynamically unstable, and a CE event ensues. Once the NS is engulfed in
the expanding envelope of the original secondary, it experiences
hyper-critical accretion (e.g., Bethe
\& Brown 1998). The spiral in of the NS toward the helium core of the
secondary ends, when the CE is expelled from the binary, at 
the expense of orbital energy. The orbital separation is greatly reduced,
and a close binary with a NS and a low-mass helium star is formed (stage
VI). The helium star evolves through subsequent burning of elements and
eventually acquires a ``giant-like'' structure, with a developed CO core
and a partially convective envelope. When it fills its Roche lobe another CE phase
begins (stage VII). However, this time, the envelope is of much lower
mass, and the binary orbit is not as greatly reduced and the first NS 
does not accrete as much mass. 
At the end of this phase the CO core of helium star is
exposed (stage VIII), and it eventually explodes as a Type Ic SN. 
Due to the very tight pre-SN
binary orbit, the probability of survival through the SN is very high, and
a very tight and highly eccentric NS-NS binary is formed with a merger
time of $\simeq 0.7$ Myr.

Group II strongly dominates the population of coalescing NS-NS systems
(87.4\%, for the standard model calculation) over group III (4.2\%) and I
(8.4\%). The formation of
Group I and II systems depends crucially on the radial evolution
(expansion) of low-mass helium stars and their response to mass loss. In
model N we do not allow for any helium-star expansion, and in model H2 we
assume that all Roche-lobe overflow events from helium-stars lead to
mergers (independent of their mass). In these models we find that NS-NS
systems are formed via classical channels (group III). The merger times of
the group III systems are typically of the order of a Gyr. 
The systems in group I and II systems are  initially much tighter
and thus short lived: their merger times are typically of the order of a Myr 
or even shorter (see Belczynski et al.\ 2002a).

\subsection{Coalescence rates}

In Table~\ref{rates} we present the coalescence rates of NS-NS binaries
for all the models we investigate. The rates have been calibrated
for the
Galactic Type II SN rates obtained by Capellaro, Evans \& Turatto (1999).  
Our standard model rate corresponds to about 50 coalescence events of
NS-NS per Galaxy per Myr. However, due to the model uncertainties, this
rate varies in a wide range of 1-300 events per Galaxy per Myr, depending
which parameter values we choose to adopt.

As shown in earlier studies (e.g., Lipunov, Postnov \& Prokhorov 1997;
Belczynski \& Bulik 1999), the coalescence rates depend strongly on the
assumed kick velocity distribution (see models B1-13). Since we assume a
rather high value for the maximum NS mass ($M_{\rm max,NS}=3.0 M_\odot$)
for the standard model, we also calculate rates for models (D1 and D2)
with lower $M_{\rm max,NS}$. The least affected subgroup
is that of the non-recycled NS-NS binaries, as they consist of low-mass
NS. However, the rates for group II and III are reduced significantly.
This is due to the specific formation channels of these two groups, in
which many compact objects are formed with masses over 1.5 and 2.0
$M_\odot$.
In models H2 and N, NS-NS are formed only in group III.
In model H2, with the lowest predicted coalescence rate of NS-NS systems, 
none of helium stars are allowed to develop convective envelopes, 
therefore we assume that any MT phase initiated by a helium star lead to 
a binary component merger.
In model N, helium stars are not allowed to expand, and therefore they never
interact, which suppresses formation of NS-NS systems in the new channels.  

All models (except C) include hyper-critical accretion onto
compact objects in CE episodes.
We find, in contrast to Bethe \& Brown (1998), that in most cases NS do not
accrete enough material to become BH, and thus the channels with phases of 
CE are still open for the formation of coalescing NS-NS binaries. 
This discrepancy is due to two facts. 
First, their assumption of $M_{\rm max,NS}=1.5 M_\odot$, which we find very 
low in view of the high NS mass estimates in Cyg X-2: $1.78 \pm 0.23 M_\odot$ 
(Orosz \& Kuulkers 1999) and in Vela X-1: $1.9^{+0.7}_{-0.5} M_\odot$ (van
Kerkwijk et al. 1995).
Second, their approximate\footnote{Bethe \& Brown (1998) have assumed that
the NS mass during CE is much smaller than the companion mass and they have 
neglected it in their eq. 5.5. However, we find this is not always true, e.g.,
see stage (VII) in our Fig.~\ref{NSevol}: with accreting NS of $1.8 
M_\odot$ and helium giant donor of only $3.1 M_\odot$.} treatment of 
accretion onto NS results in higher final compact object masses than in our 
exact numerical solution.
Both, our detailed treatment of the hyper-critical accretion and the
variation of coalescence rates of compact object binaries for all our 
models is given in Belczynski et al. (2002a).

\subsection{Distribution of merger sites}

\begin{figure*}[t]
\centerline{\psfig{file=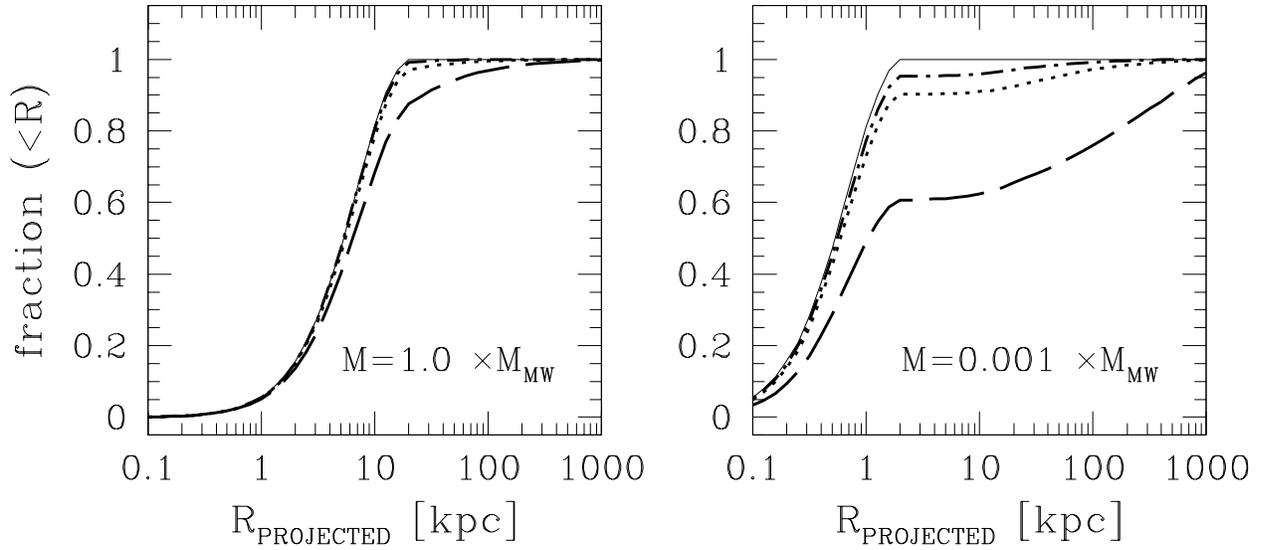,height=0.5\textwidth}}
\caption{ 
Standard Model. 
Cumulative distributions of merger sites  of
coalescing NS-NS binaries, around a massive galaxy with $M=M_{\rm MW}$ (left
panel) and dwarf galaxy with $M=0.001 \times M_{\rm MW}$ (right
panel); group I:
thick dotted line, group II: thick dot-dashed line, and group
III: thick long dashed line.  and the initial distribution of primordial 
binary population within the given mass galaxy is given with the solid thin 
line.
}
\label{newtwo}
\end{figure*}

We first consider the distributions of merger sites  for each
Group identified above. In the left panel of
Figure~\ref{newtwo} we present the case of a massive Milky Way like
galaxy. 
The distribution of merger sites of groups I and II follows
closely the initial stellar distribution in this galaxy. This is expected
since these systems are very tight, and therefore their lifetimes as
NS-NS systems are  typically  a few
Myr. Even with a velocity of a $1000$km\,s$^{-1}$ a NS-NS 
system will move only out to $\sim 1$\,kpc before it merges. This is also
seen in the right panel of Figure~\ref{newtwo} where we show the
distributions around a small mass host. The binaries of group I and II
merge within such galaxies, and the distribution of the merger sites
follows closely the initial distribution of stars. The classical systems,
(Group III) have much longer merger times and therefore some of
them manage to escape even from high mass hosts,  see
Figure~\ref{newtwo}. In the case of small galaxy the distribution of
merger sites of group III binaries contains a large fraction that merges
far outside the host. One must bear in mind that our simulations show that
the entire population of NS-NS systems is dominated by 
group II systems, and group III systems are the smallest subgroup. 

\begin{figure*}[t]
\centerline{\psfig{file=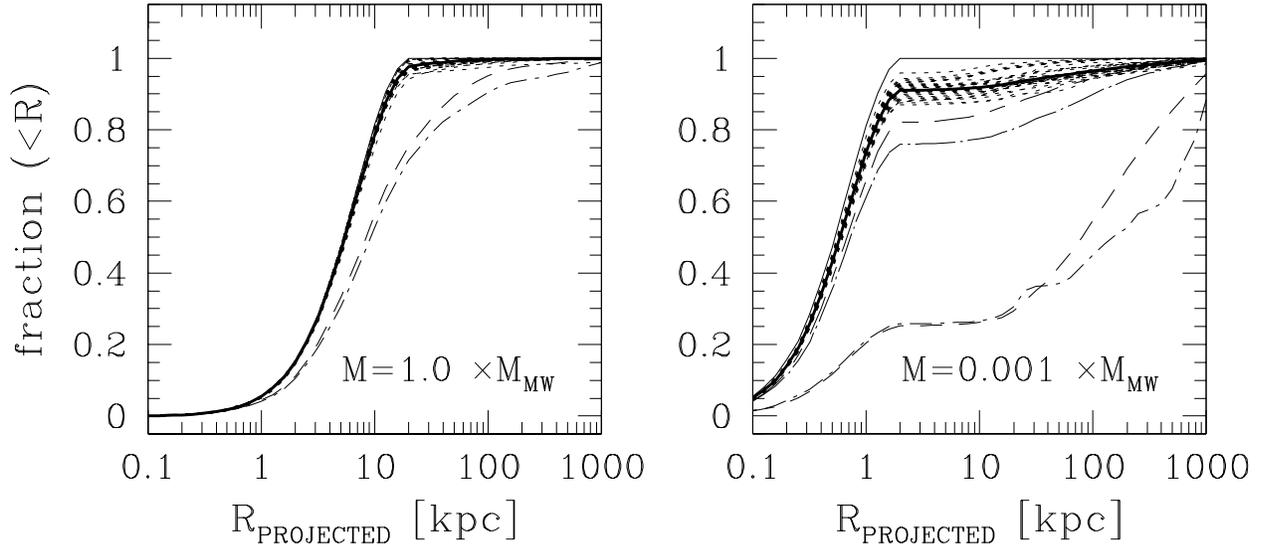,height=0.5\textwidth}}
\caption{ 
Parameter Study. Cumulative distributions of NS-NS merger sites for
different evolutionary models.
The left panel shows distribution for NS-NS systems born in a massive
galaxy ($M=1.0 \times M_{\rm MW}$) and the right panel for a dwarf 
galaxy ($M=0.001 \times M_{\rm MW}$).
Initial distribution of primordial binary population within the given
mass galaxy is plotted with the thin solid line, while different model
distributions are plotted with dashed and dotted lines.
Three most extreme distributions are marked: with short-long dashed line
that of model E3, with dotted long dashed line that of model F2, 
dotted short dashed line that of model H2 and with short dashed line that 
of (nonphysical) model N.
Standard model distribution is marked with thick solid line.
}
\label{newthree}
\end{figure*}

We present the distributions of NS-NS merger sites around the two model 
galaxies for all our evolutionary models listed in
Table~\ref{rates} 
in  Figure~\ref{newthree}.
In the case of a massive host galaxy, all distributions  
except for models H2 and N nearly overlap. The escaping fraction,
defined as numbers of systems merging further than 20~kpc from the host is
always smaller than five percent. In models H2 and N formation of group I and
group II systems is suppressed and the escaping fraction is much larger.
In the case of a dwarf host (right panel of Figure~\ref{newthree}) the
escaping fraction varies between 5 and 15\%. For models E3 ad F2 the
distributions clearly stand out, yet even in these cases nearly 80\% of the
mergers take place within the host galaxies. The E3 systems are formed
assuming a very high CE efficiency, and therefore are still wide after the
CE phase.  Thus they are relatively long lived and this broadens the
distribution of merger sites.
In model F2 every MT event (except a CE
phase) is treated conservatively, i.e., all material lost from the donor
is accreted by the companion. Therefore the separation following MT events 
is wider, as no material and thus no angular momentum is lost from the binary. 
Wider systems live longer and therefore may merge further from the host.
Models H2 and N, roughly represent  distributions of NS-NS binaries
formed only through classical channels with long merger times, that merge 
far from their birth places (for small hosts the escaping fraction is 
$\sim$\,70\%).

\section{Conclusions}

Previous calculations of  distribution of merger sites of NS-NS
binaries showed 
that NS-NS binaries merge predominantly outside of host
galaxies. This result was used as an argument against the  NS-NS
systems as GRB progenitors.   With the calculations presented
here we show that the NS-NS population may consist primarily of
systems with small orbital separations  and consequently short
merger times, which merge close to the  place where they had
been  formed. Such systems are  formed along  the newly
identified formation
channels, opened in our  calculations for the two following
reasons.   The first is  our treatment of the evolution of low-mass 
helium stars, which are allowed to form partially or fully
convective envelopes. The second is that  we have considered 
double common envelope episodes of two giant stars  (H- and He-rich) 
leading to ejection of  their combined  envelopes. These
are the two  major features of our calculations that allow the
formation of  the two  new subpopulations of NS-NS binaries.
Evolution of helium stars in CE phases has been considered 
qualitatively in earlier studies (van den Heuvel 1992; Taam 1996).
MT interactions involving helium stars (although not CE phases) 
have been included in population synthesis study of Tutukov \& 
Yungelson (1993).

If NS-NS mergers are the progenitors of short bursts then we find
that they should lie within galaxies just as the long bursts 
do. We note that the BH-NS mergers  take place
preferentially outside galaxies (for details see 
Belczynski et al.\ 2002b).  Current and future space missions
will hopefully measure precise positions of short bursts and
settle the issue of their progenitors.

\acknowledgements We would like to thank the referee D.\ Chernoff and also 
C.\ Fryer, R.\ Taam, F.\ Rasio, W.\ Bednarek, and J.\ Zio{\l}kowski for comments 
on this work.  
Partial support is acknowledged from NSF Grant PHY-0121420 (VK),  
from KBN Grant 5P03D01120 (TB; KB) and the Aspen Center for
Physics (TB; VK).

\pagebreak

\begin{deluxetable}{lrrrrp{5cm}}
\tablewidth{280pt}
\tablecaption{ Galactic NS-NS Coalescence Rates (Myr$^{-1}$)}
\tablehead{ Model&  I&  II&  III& Total &Model details\tablenotemark{a} } 
\startdata

A    &  4.4&  46.1&  2.2&  52.7 & standard model \\   
B1   &  6.0& 286.4&  0.0& 292.4 & zero kicks \\ 
B3   &  6.4& 295.7&  0.1& 302.2  &$\sigma=20$\,km\,s$^{-1}$\\
B6   &  6.5& 219.8&  0.6& 226.8  &$\sigma=50$\,km\,s$^{-1}$\\
B7   &  5.2& 121.0&  1.9& 128.1  &$\sigma=100$\,km\,s$^{-1}$\\
B9   &  4.3&  27.5&  1.4&  33.2  &$\sigma=300$\,km\,s$^{-1}$\\ 
B12  &  1.9&   5.8&  0.3&   8.0  &$\sigma=600$\,km\,s$^{-1}$\\ 
B13  &  4.6&  84.1&  2.3&  91.0  & ``Paczynski'' kicks \\ 
C    &  3.2&  37.5&  2.6&  43.2  & no HCA in CEs \\
D1   &  4.9&  28.0&  0.7&  33.6  &  $M_{\rm max,NS}=2$\,M$_\odot$ \\ 
D2   &  3.6&   5.5&  0.0&   9.1  & $M_{\rm max,NS}=1.5$\,M$_\odot$\\
E1   &  0.4&   2.0&  0.3&   2.7  & $\alpha_{\rm CE}\times\lambda = 0.1$\\  
E2   &  3.1&  19.0&  1.4&  23.5  & $\alpha_{\rm CE}\times\lambda = 0.5$\\
E3   &  5.2&  99.8&  4.0& 109.0  & $\alpha_{\rm CE}\times\lambda = 2$\\  
F1   &  2.3&  18.6&  1.2&  22.1  &  f$_{\rm a}=0.1$\\
F2   &  2.3&  44.6&  7.5&  54.3  &  f$_{\rm a}= 1$\\
G1   &  3.3&  38.7&  1.9&  43.9  &wind decreased by 2\\ 
G2   &  7.3&  82.7&  2.2&  92.2  &wind increased by 2\\ 
H1   &  3.3&  33.0&  1.6&  37.9  & $M_{\rm conv}=4.0$$\,M_\odot$\\ 
H2   &  0.0&   0.0&  0.9&   0.9  & $M_{\rm conv}=0$$\,M_\odot$\\ 
I    &  4.0&  47.4&  3.2&  54.5  &  burst-like SFR\\ 
J    &  4.4&  50.7&  3.0&  58.1  &  (IMF):$\propto M_1^{-2.35}$\\
K1   &  1.8&  19.7&  1.0&  22.5  & binary fraction: 25\% \\
K2   &  7.4&  79.0&  3.8&  90.2  & binary fraction: 75\% \\
L1   &  6.3&  66.1&  6.5&  78.9  & $j_{\rm a}=0.5$\\  
L2   &  2.0&   9.6&  0.4&  12.0  & $j_{\rm a}=2.0$\\
M1   &  0.2&   5.8&  0.2&   6.2  & $\Phi(q) \propto q^{-2.7}$\\
M2   & 14.0&  94.3&  5.9& 114.2  & $\Phi(q) \propto  q^{3}$\\ 
N    &  0.0&   0.0& 34.4&  34.4  & no He-star expan.\\  
O    &  4.3&  45.0&  2.6&  51.9  & extended FB\\ 
\enddata
\label{rates}
\tablenotetext{a}{For definition of parameters see \S\,2}
\end{deluxetable}

\end{document}